# Super-forecasting the 'technological singularity' risks from artificial intelligence

Petar Radanliev[1], David De Roure[1], Carsten Maple[2], Uchenna Ani[3]

[1]Oxford e-Research Centre, Department of Engineering Sciences, University of Oxford, UK, petar.radanliev@oerc.ox.ac.uk; [2]WMG Cyber Security Centre, University of Warwick, [3]STEaPP, Faculty of Engineering Science, University College London

**ABSTRACT**

This article investigates cybersecurity (and risk) in the context of 'technological singularity' from artificial intelligence. The investigation constructs multiple risk forecasts that are synthesised in a new framework for counteracting risks from AI itself. In other words, the research concern in this article is not just with securing a system, but to analyse how the system responds when (internally and externally caused) failure and compromise occur. This is an important methodological principle because not all systems can be secured, and we need to construct algorithms that will enable systems to continue operating even when parts of the system have been compromised. Furthermore, the article forecasts emerging cyber-risks from the integration of artificial intelligence in cybersecurity. Based on the forecasts, the article is concentrated on creating synergies between the existing literature, the data sources identified in the survey and forecasts. The forecasts are used to increase the feasibility of the overall research and enable the development of novel methodology that uses AI to defend from cyber risk. The methodology is focused on addressing the risk of AI attack, as well as to forecast the value of AI in defence and in the prevention of AI rogue devices acting independently.

## 1   Introduction

Artificial intelligence (AI) can be described as an autonomous and self-evolving system that can recognise and learn from unknown and unpredictable data patterns. AI systems can continuously evolve and learn and improve their domain adaptation and self-organization - after being designed. While this creates many opportunities for self-improving evolving systems, it also creates risks from such systems being used by adversaries against its original intentions. There is a growing concern caused by the increased adoption of artificial intelligence (AI) in predictive cybersecurity, triggering various discussions on the 'Skynet' becoming a reality. These fears are amplified by public figures (e.g., Stephen Hawkins) and tech gurus (e.g., Elon Musk) arguing that AI is a serious risk to humanity and could result in human extinction. We agree that if AI continues to be used in defence and security, could eventually lead to a 'technological singularity' event, causing unpredictable changes to human civilization 'that would signal the end of the human era' [1]. The aim of this article is to create complementarities between the topics of AI and cybersecurity, to promote adaptation (i.e., focusing on trust in AI systems) and to enable the categorisations of risks (which is necessary for quantifying the cascading effects of cyber-risk). Since the rise of AI seems inevitable, the objectives of this article are to forecast areas that we need to address to mitigate the probability of a 'technological singularity' event, not to prevent it from occurring, because following the existing speculation models, that presumption seems inevitable [2].

The study initiates with a brief literature review, followed by a survey of secondary data sources, upon which the initial results and forecasts are grounded. The final section of this article is addressing the most important challenge in the development and application of novel AI algorithms for the application of AI in cybersecurity; the cyber-risks from AI itself. In other words. The final chapter of this article analyses and forecasts cyber risk from rogue AI systems.

## 1.1 Trends in artificial intelligence and cybersecurity

We used Google Trends to compare the search interest on artificial intelligence (AI), cyber security and cyber risk, and we analysed the trends on these topics over the time period from 2004 to 2021. The numbers in Figure 1 represent search interest 'relative to the highest point on the chart for the given region and time' where 100 represents the peak popularity, 50 represents that the half as popular, and 0 represents a lack of data for this term i.e., lack of popularity and interests.

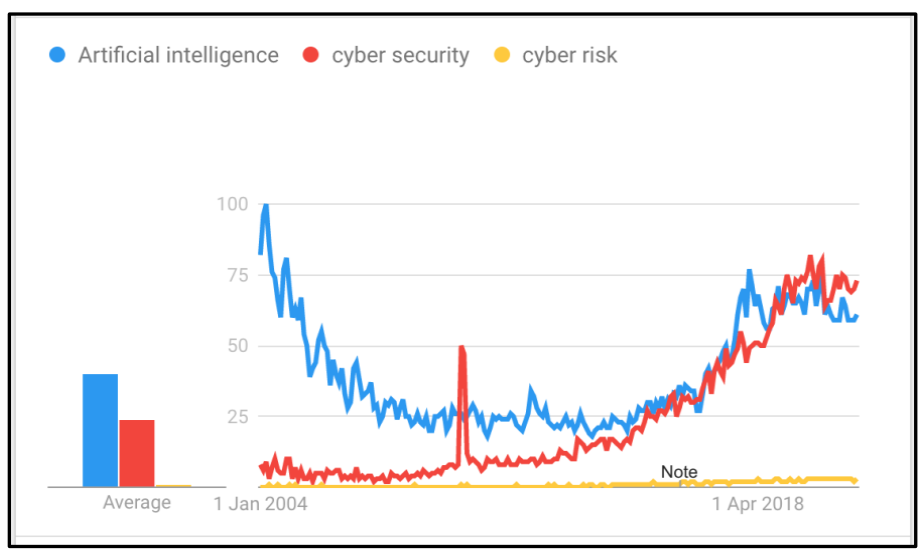

*Figure 1: Search trends on the topics of AI, Cybersecurity and Cyber risk*

From Figure 1, we can see that AI has been a more popular term in search trends, but since 2018, the cybersecurity has become a more interesting term, with a score of 73, comparing with the AI score of 61, and cyber risk score of only 3 – data recorded in February 2021. We set the search parameters to 'worldwide', because we wanted to analyse the global trends. If we change the parameters to a specific region, the results also change. But the low interest in the topic of cyber risk is consistent across regions. It seems that the world is more interested in securing the cyber world, than the risks from the increasingly connected world.

## 2  Methodology

AI has traditionally been used for defence, to prevent intrusion and cyber-attacks, but sooner or later, the intruders will start using AI for cyber-attacks. Current defence mechanisms would limit the AI attack to specific segments, but what if AI attack is successful, what would prevent the attacker accessing proprietary information? One way to prevent this is with a multi-layer network defence, where AI would enforce cryptography - when attacker reaches a certain level. Then, to identify the level of the attacker in a multi-layer network defence, a Network Intrusion Detection System (IDS) can be used. In such example, a User Behaviour Analytics can be used - to observe users and devices behaviour. Such approach would help with resolving the big data problem when analysing large networks. However, there are many gaps remaining in applying this approach in cloud security and IoT security.

To identify how cybersecurity can be enhanced with AI, we initiate the study with a literature review and present an overview on how attackers use AI and Machine Learning (ML) to attack IoT systems. Since ML (and the current AI) are simply algorithms, the overview is used to build a categorisation of innovative design concepts. The categorisations can be used to build robust IoT systems, that are intrinsically secured for cyber-attacks. In the review and categorisation, we considered AI as a concept, and ML as an approach. Thus, most of cyber security is related to ML (approach) and not AI (concept). Building upon this description, we should clarify that ML and AI are based on using algorithms, while statistical (quantitative) risk assessment is based on using data analytics. The fundamental difference between statistical estimation and ML is that statistical estimation uses linear techniques. However, ML can be aligned for calculation or prediction, for example by using evolutionary algorithms, and ML can present more accurate estimations.

Since most of the cyber security tasks are automated and not human related, then we are considering ML as a subset of AI, in a similar fashion as applied in many real time systems e.g., navigation systems, radars, satellites. With this approach, AI algorithms can predict cyber risk dynamically and serve as an early alert system - even if the forecasting is based on the use of judgement as input. In terms of quality of the early alert/detection system, we consider the: identification, definition, signs, and preconditions of risk types, as the main factors that determine quality of the dynamic early detection system. In this evaluation, the cyber risk assessment should be conducted on the data sets describing the preconditions - prior the detection.

Full reference:
Radanliev, Petar., De Roure, David., Maple, Carsten., and Ani, Uchenna, "Super-forecasting the 'technological singularity' risks from artificial intelligence," *Evol. Syst. 2022*, pp. 1–11, Jun. 2022, doi: 10.1007/S12530-022-09431-7, URL: https://link.springer.com/article/10.1007/s12530-022-09431-7.

## 3 Literature review

The topics of AI and cybersecurity are constantly advancing, and what was considered as state of the art five years ago, is in many cases obsolete today. On the other hand, there are some established principles that have not changed in a long time. For example, every AI algorithm that we use today, is based on one algorithm that was developed 34 years ago [3]. To tackle this disparity in relevance, firstly a short literature review is conducted on the most prominent literature on the topics of AI and cybersecurity. Secondly, a short literature review is conducted on the most recent studies on the same topics. In the search for literature, three different search engines were used, (1) Web of Science; (2) Scopus; and (3) Google Scholar. For the first part, citations and relevance were used as parameters. For the second part, time and quality of the journal were used as reference points. We adjusted the parameters, because in the second search, given the short time from publication to present date, the citations might not reflex the value of the research paper. We considered the ranking of the journal as more suitable reference point in the pursuit of the most valuable recent literature.

### 3.1 Current state of the art

At present, AI application in cybersecurity is predominately focused on using big volumes of data for discovering attack changes and flexibility in response to threats [4]. With the increased applications of AI in smart cities design, the role of AI is increasingly dominating in smart grids, intelligent transport systems, and autonomous vehicles [5]. For example, evolving ANN-based sensors are already used for processing, prediction and anomaly detection in context-aware cyber physical systems [6]. Cyber-attack predictions have been produced by generating attack graphs and predicting future attacks, and proven both practical and effective [7]. Similar methods have been adapted for distributed anomaly detection in IoT, with the graph neural network method [8]. In Table 1 we present and review the current state-of-the-art in applications of AI in cybersecurity.

*Table 1: Current state of the art in AI - cybersecurity*

| Research topics | Methods | Data interpretation: qualitative or quantitative | Arguments and Contribution | Reference |
|---|---|---|---|---|
| AI, cybersecurity | Literature review | Qualitative | Swarm behaviour patterns can be incorporated in antimalware systems | [4] |
| AI, smart cities | Literature survey | Qualitative | Review of AI techniques | [5] |
| AI, social media, cybersecurity | Review | Qualitative | AI can be integrated in social media cybersecurity | [9] |
| AI, cybersecurity | Review | Qualitative | Description of cyberattacks on network stacks and applications | [10] |
| AI, cybersecurity | Review | Qualitative | AI has facilitated a reduced model training time | [11] |

From the review of different methods (in Table 1) used for investigating the topic of AI in cybersecurity, we can see that literature review and different types of qualitative reviews are predominating the current literature on data interpretation. This is applicable to various arguments and contributions, starting from swarm behaviours in antimalware systems and cyberattacks on network stacks, to AI in social media cybersecurity. In the next section, we discuss the availability of data sources for conducting a quantitative cyber risk assessment.

Full reference:
Radanliev, Petar., De Roure, David., Maple, Carsten., and Ani, Uchenna, "Super-forecasting the 'technological singularity' risks from artificial intelligence," *Evol. Syst. 2022*, pp. 1–11, Jun. 2022, doi: 10.1007/S12530-022-09431-7, URL: https://link.springer.com/article/10.1007/s12530-022-09431-7.

# 4 Survey of data sources

## 4.1 Data Sources for quantitative cyber risk assessment

One of the main difficulties in quantitative cyber risk assessment is the lack of probabilistic data. It is not that such data does not exist, it is more that data strategies for collecting such data do not exist. Given the lack of standards and regulations on cyber risk data strategies, organisations need to adapt their data collection according to their risk assessment requirements. In Table 2 we list some of the potential data sources, that when combined, can provide a significant (data-rich) improvement on the cyber risk quantification problem.

*Table 2: Examples of cybersecurity data sources*

| Type | Platforms | Metadata | Data source | Description |
|---|---|---|---|---|
| External | Shodan, Censys, Fofsa, BinaryEdge | IP, banner data, images | IoT search engines | Search engines of publicly accessible IoT devices |
|  | Hansa, DreamMarket | Product/author name, price | DarkNet marketplaces | Markets for illicit goods |
|  | EMBER, VirusTotal | Hash, binary, date, malware reports | Malware Repositories | Sites collecting malware reports |
| Internal | File store, disk drives, file directories | File size, directory name, file name, directory size | File store, disk drives, file directories | Devices that store data from users and networks |
|  | BurpSuite, Nessus, Qualys, OpenVAS | Name, severity, risk | Vulnerability assessment | Reports from vulnerability scanning tools |
|  | Docker, containers, VMware | Operating system, applications, file systems | Workstations and virtual machines | Computational machines |

These data sources (along with many other data sources) have already been applied in different cybersecurity solutions – see Table 3. These solutions have proven effective in identifying new threats (cyber threat intelligence), and various tools have been designed for advanced phishing, dynamic and static analytics, and many more operational solutions. One of the solutions that currently predominates in news media is the application of AI algorithms for disinformation and computational propaganda. In recent years, the fake news and online propaganda has proven to be a significant threat that can destabilise governments even in the most developed and secured countries i.e., USA. The small scale (qualitative) methods for fake news and disinformation have always existed. But with the rise of social media giants (e.g., Facebook, Twitter), and the emergence of various new social media platforms (e.g., Reddit, Weibo), adversaries have stared deploying AI algorithms to create targeted computational propaganda and fake news / disinformation strategies. AI algorithms can process different types of big data (e.g., video, text, audio, images) with techniques such as text mining, image recognition, and apply for AI driven cyber-attacks (e.g., astroturfing, bots, message amplification).

Full reference:
Radanliev, Petar., De Roure, David., Maple, Carsten., and Ani, Uchenna, "Super-forecasting the 'technological singularity' risks from artificial intelligence," *Evol. Syst. 2022*, pp. 1–11, Jun. 2022, doi: 10.1007/S12530-022-09431-7, URL: https://link.springer.com/article/10.1007/s12530-022-09431-7.

*Table 3: Examples of AI applied in cybersecurity*

| Application | Tasks | Datasets | Tools | Companies |
|---|---|---|---|---|
| Security Operations Centres | Log file analysis | Boss of the SOC | Kiwi, Splunk | Splunk |
| | Vulnerability assessment | National Vulnerability Database, Metasploit | Nessus, ZMap | Tenable |
| | Intrusion detection | CIC-IDS 2017 | Zeek | Palo Alto |
| Disinformation/ Computational Propaganda | Bot detection | Bot Repo, Twitter Bot-Cyborg | Hoaxy, Botometer | Paragon Science |
| | Disinformation identification | Credibility Coalition, Grand Old Party, Twitter | Exifdata, exiftool, factcheck | CarleyTech, Rand, FireEye |
| Cyber threat intelligence | Malware analysis | VirusTotal | Cuckoo | FireEye |
| | Phishing detection | PhishTank | PhishMonger | KnowBe4 |
| | Dark Web Analysis | AZSecure HAP | ISILinux | CYR3CON |
| Adversarial ML | Malware evasion, ML poisoning | EMBER, Neural Information Processing Systems Adv. learning | EvadeML, SecML | Elastic, Google Brain, Microsoft |

Traditionally, the security and operations centres are at the core of human-cantered cybersecurity, but this approach has created many false positives and negatives, and such human-centred systems have also exhibited ease of overloading. AI has started to emerge as a more capable solution for filtering (big data) results. But AI has also been used by adversaries to trick the defence algorithms by including deceptive and polluted data (e.g., deep fakes, synthetic text, video, images). Most concerning developments are the tools and techniques developed in the areas of generative adversarial networks, reinforcement learning, and actor critic networks. By using these tools and techniques, adversaries can teach algorithms to evolve in a dynamic environment, and mimic the human learning process, with a limited training data.

## 5   Results of the Super-Forecasting

### 5.1   Super-Forecasts

AI algorithms have already resolved several cyber-security problems, such as automatic behaviour analysis, human-computer interaction patterns analysis, and the design of intelligent anti-virus software.

### 5.1.1 Current AI cyber-risk solutions include:

1. Network Intrusion Detection System (IDS): can be used to classify the network behaviour - of the user as normal behaviour or cyber-attack. One example of IDS enhancing the classification accuracy of the anomaly detection, is by applying machine learning models (i.e., classifiers) such as Artificial Neural Network (ANN), Support Vector Machine (SVM), Decision Tree (DT), or Naive Bayesian Classifier (NBC). In such example, classifiers can be enhanced by using metaheuristics, e.g., an algorithm used to train ANN (i.e., finding the weights of the nodes), or finding the optimal parameters for SVM. Another example is the 'Feature Selection Problem' where AI id used to enhance IDS, by selecting the most significant features in the IDS dataset (i.e., NSL or KDD dataset). This (feature selection) relies on the best approach to select a minimum number of features, that would enhance classification accuracy of the algorithm, or: $Min^{\{No.of\ Features\}} = Max^{\{Classification\ Accuracy\}}$
2. AI in email or text/image messages spam filtering and/or malware detection: AI can classify emails and/or text/image messages into spam & not spam. In such example, AI and ML algorithms could be applied as in the previous example.

### 5.1.2 Forecasts on how AI will improve cyber-risk assessment in near future:

1. With the emergence of Covid-19 more people started working from home. This has changed the cyber security and requires a virtual security of fast-growing endpoint connections. In near future, AI will be used for large scale - fully remote lifecycle management of IT devices. As early as 2021, organisations will focus on securing the end point devices. This would ensure secure remote access to services, data and resources, regardless of whether the access point is in the office or at home. Hence, as a result of Covid-19, the cybersecurity would emerge stronger, and AI will be the leading force in innovations of remote cybersecurity.
2. The emergence of Covid-19 has amplified the existing shortage of cyber security experts. Intruders are already exploiting on the confusions triggered from the new home working arrangements, and this has resulted with one of the largest hacks in 2020 on the US government. In near future, we can forecast the increased use of AI in defence, starting with a large-scale deployment of AI for recognising patterns of attacks, suspicious activity monitoring, and large-scale automation of defence against phishing, ransomware, etc. Such large-scale deployment of AI innovations in cyber defence, would free up the existing security experts to perform more hands-on security tasks.
3. Since Covid-19 has triggered a massive shift to home/office working, we can expect attackers to capitalise on this change. AI bots will be used, in combination with social engineering techniques and there are millions of bots currently live on the internet. This will trigger a new deployment of AI to determine malicious bots.
4. We can forecast a rise of artificial intelligence for IT operations (AIOps) through multi-layered tech platforms for continuous integration and deployment (CI/CD) automating and enhancing IT operations though ML analytics. The rise of AIOps will occur in 2021, because the ITOps complexity management with the traditional human interventions is no longer an option. ITOps have exceeded the human scale and Covid-19 just made that more obvious. The increasing new and emerging forms of data (e.g. from IoT devices, APIs, mobile apps, etc.), combined with the difficulties caused by Covid-19, would simply become too complex to resolve without AIOps. Additional reasons for forecasting the rise of AIOps in 2021, is the increased commercialisation and user/business dependence on IT infrastructure due to Covid-19, which has changed the users and industry expectations for cyber-attacks and IT events causing infrastructure problems, to be fixed at increasing speeds.

5. In 2021, ITOps will start shifting from core IT functions, to the edge of the network. The advancements in cloud infrastructure and third-party services, will result with IT budgets being relocated from core IT functions to the edge, where additional computing power can be added on request. This will result with more monitoring responsibility being forced upon developers at the application level, while the overall accountability would remain a core IT function. This will result with ITOps accepting more responsibilities, while the networks would continue to become more complex. To cope with this increasing complexity, the ITOps function would have to evolve into AIOps.

6. In 2021, the data strategies will evolve to support the evolution of AI. One example of how data strategies will change in 2021 is the integration of extensive and diverse IT data (e.g., metrics and events data from IT operations management will be visible along with incidents and changes data from IT services management). The integration of these diverse data sets would enable automation of cyber-risk analytics. This data integration will result with a vast amount of diverse data that won't be possible to analyse with manual effort. Therefore, this integration of real-time big data would result with a platform that would support real-time cyber analytics with ML.

### 5.1.3 Forecasts on how AI will be used for cyber-attacks in near future:

1. In the same way AI can monitor the network to detect cyber-attacks, adversaries will use AI to observe cybersecurity defensive decisions and use Deep Learning (DL) network for automatic adversarial attacks against ML defence systems.
2. AI will be used to generate poisoning attacks, targeted at defence AI systems and poisoning the training data, resulting with inaccurate cyber-defence outputs based on the polluted learning data.
3. Since most of the training data for AI defence algorithms are based on public and open access records on data breaches, adversaries will use the same data, or conduct training data stealing to learn how defence algorithms operate and design AI that can surpass the defence engines.
4. Adversarial AI will create false positive and false negative misclassifications to disguise an actual attack.

### 5.1.4 Forecasting how and why cybersecurity will fail as a result of AI

1. New cybersecurity solutions based on AI algorithms are developed as isolated systems in industry and in academia. Academic researchers use old datasets to

   develop very specialised algorithmic solutions, which perform with great precision in testing environments, but fail in operational environments, because the threats have evolved since the data was collected. With industry, the reason for failure is the complete opposite, organisations have vast amounts of data, but use old algorithms, which are not trained to detect new and emerging cyber-risks.
2. Considering the first forecast – that academia is leading the research on AI solutions for cybersecurity – the performance analysis is conducted mostly on single data sets. These solutions would have been much stronger if the performance analysis is tested on multiple data sets simultaneously, which can only happen in industry setting – because of data availability. Adversarial AI algorithms are not bound by ethical considerations and data privacy regulations. Since these data privacy regulations are only applicable to the training data in defence algorithms, and if alternatives are not found very soon, the defence algorithms will likely lose to adversarial algorithms – that are trained for practical applications.
3. Defence algorithms that are designed as very specialised solutions, are almost never deployed and used for other purposes and lack transparency. This lack of model sharing and lack of interoperability results with a lot of work and effort invested in a defence algorithm that will probably never be adapted and reused for a different purpose. The key for continuous improvement is in sharing information and details on the type of algorithm, this will enable existing algorithms to be adapted and quickly and timely developed into new and relevant algorithms that can be deployed in a different (across) environment. Since adversarial algorithms are commonly shared (for a price), it is likely that adversarial algorithms will have the upper hand in this area of interoperability and applications across different environments.
4. The adoption of AI solutions for cybersecurity is still very costly, and almost unaffordable for small and medium sized organisations. These solutions are highly specialised and used to perform specific functions at speed and at scale. Hence, even if the cost were reduced (e.g., direct subsidies), the solutions might not be that relevant to small and medium sized companies. This would create advantages for adversaries, who could target these companies at scale – with AI driven cyber-attacks.

## 5.2 Solutions for enhancing cybersecurity with AI

Cybersecurity solutions based on AI algorithms should apply multi-view / multi-modal analytics, using multiple data sources in deep learning based approach e.g., deep structured semantic models, followed by multi source approach, and multi-task learning strategies. Such integrated approach can produce improved risk management and a better understanding of cyber risk maturity and security posture. A crucial cybersecurity advantage of deep learning methods is the ability to detect hidden patterns in the data, which can be used to detect a zero-day threat. One of the crucial weaknesses of such integrated deep-learning approach to cybersecurity, is that it presents a 'black-box' and experts cannot explain how the algorithm reaches decisions. Since the model cannot be explained this affects its trustworthiness and makes it really hard to reuse it and make it interoperable. Future solutions need to be based on opening up the 'black-box', and making algorithms easy to understand and explain, and make them interoperable. This can be done by using 'post hoc' and 'intrinsic' algorithms. 'Post hoc' algorithms can explain (interpret) their training and examine the training components and processed – after been trained. Intrinsic algorithms in a similar way can interpret the type of data that was used to reach a decision. These approaches must be used, so we can understand and explain how algorithms learn, and make them interoperable. The explaining and understanding process of the main characteristics from such big datasets requires visualisation mechanisms. Depending on the data, such visualisation can be expressed as simple charts and tables, or more complex geo-spatial layouts with different colours. Designing such visualisations, with various options as an overview with zooming, filtering, and many different options, can enhance significantly the efficiency and speed of cybersecurity deployment. Visualisations in combinations with predictive analytics would present a significant advancement in the current state of the art in cybersecurity. AI methods that can improve the current predictive analytics include temporal-based graph neural networks, burst detection, deep generative modelling with temporal constraints, and deep Bayesian forecasting.

## 6   Forecasting the required solutions for cyber-risks from AI itself

This section **constructs** solutions for an alternative design of AI in cybersecurity (Figure 2), in which the **systems continue to operate**, without compromising the security and privacy of the critical systems.

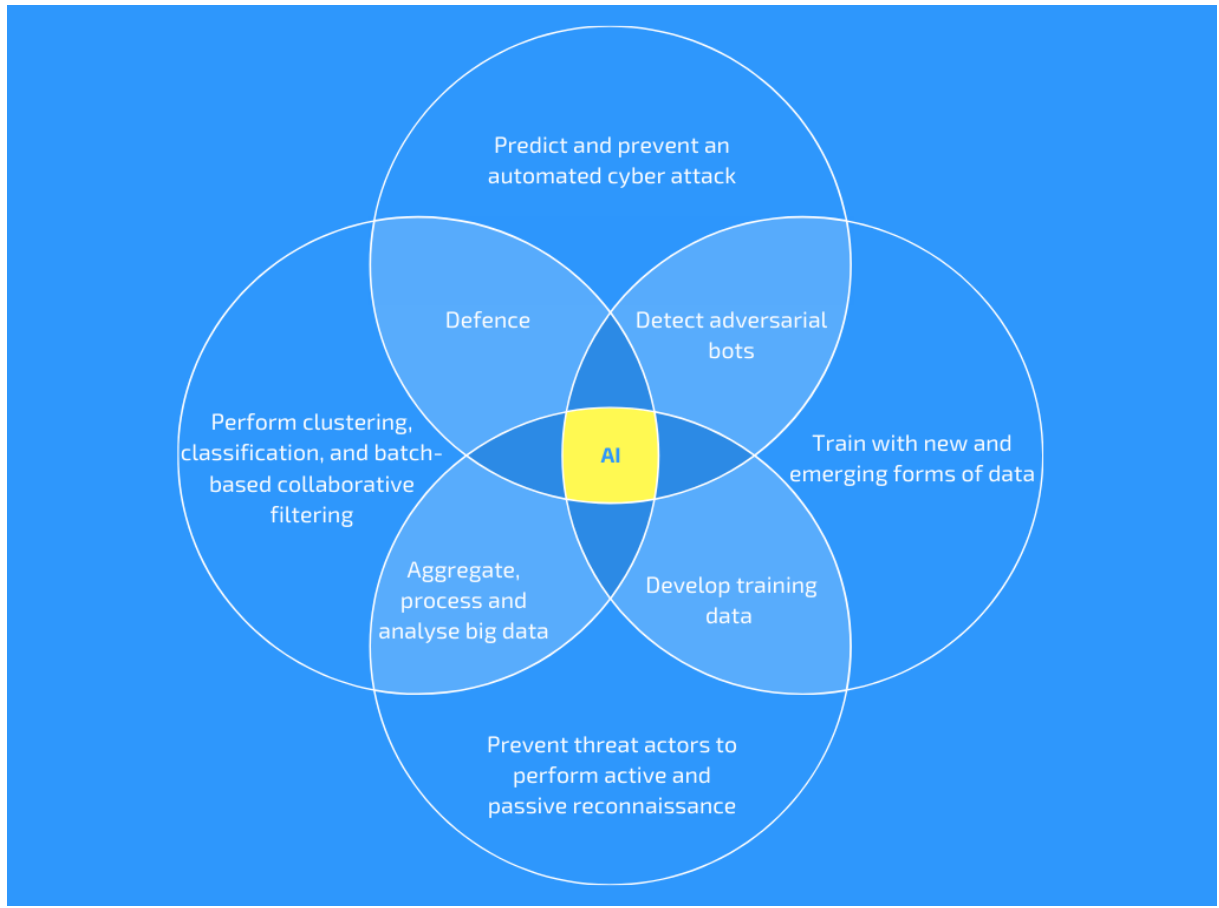

*Figure 2: Alternative design of AI in cybersecurity*

To achieve this, we need algorithms (A) that can classify how AI-driven bot can analyse big data to predict and prevent an **automated attacs**. For the first algorithm ($A_1$) we need to construct training scenarios that will teach AI to use the OSINT (Big Data) to predict and prevent an automated cyber-attack. Then use the scenarios with modern tools such as Recon-ng, Maltego, TheHarvester, Buscador, OSINT sources - to develop and train a **new transferable AI algorithm** for cyber **defence** in other sectors (e.g., finance) as a preventative solution.

For the second algorithm ($A_2$) we need to map how adversarial AI could pollute the training data in a way that seems legitimate (e.g., using direct references to results obtained from OSINT queries). We need to identify how to train the algorithm to learn patterns of data pollution and become more efficient for cyber defence. This would involve constructing a scenario to **teach the algorithm how adversarial systems operate** and how to **build systems** that will **prevent** such scenarios from happening. To identify training data for the second algorithm we need to expand the search in **new and emerging forms of data**, e.g., **open data** – Open Data Institute[1], Elgin[2], DataViva[3]; **spatiotemporal data** - GeoBrick [12], Urban Flow prediction [13], Air quality [14], GIS platform [15]; **high-dimensional data** – Industrial big data [16], IGA-ELM [17], MDS [18], TMAP [19]; **time-stamped data** – Qubit[4], Edge MWN [20], Mobi-IoST [21], Edge DHT analytics [22]; **real-time data** – CUSUM [23], and **big data** [24].

---

[1] https://theodi.org/
[2] https://www.elgintech.com/
[3] http://dataviva.info/en/

Thirdly, we need to construct an algorithm ($A_3$) that can map the future cyber-risks and identify how adversarial AI can crawl web sites, DNS records, and many OSINT sources to build a profile of the target. **Then develop the training data** for a new AI algorithm aimed at **detecting** cyber-attacks at the **edge of the network in real time and** synthesise recommendations on training data for detecting **non-technical cyber-attacks**, e.g., social engineering attack.

We can use the training data to **build and train** a new AI algorithm ($A_4$) that can **prevent threat actors** to perform **active and passive reconnaissance** about a specific target at scale. Then we need to teach the new AI algorithm how to (1) use Samsara to write it's own improved algorithms (2) by using Spark with its machine learning library, (3) use MLLib for iterative machine learning applications in-memory, (4) use MLLib for classification and regression, and to build machine-learning pipelines with hyper-parameter tuning.

We also need to teach a new algorithm ($A_5$) how adversaries use Spark to aggregate, process and analyse the OSINT big data and to process data in RAM using Resilient Distributed Dataset (RDD). Then teach the algorithm how to use Spark Core for scheduling, optimisations, RDD abstraction, and to connect to the correct filesystem (e.g., HDFS, S3, RDBMs). We also need to make improvements to the AI algorithm based on the results after training the algorithm for cyber risk analytics in real-time.

We need to train a new algorithm ($A_6$) how to use data sources from existing libraries such as MLLib for machine learning and GraphX for graph problems. This can be applied to categorise how a defence algorithm can identify adversarial techniques and bots efficiently and with low cost. Use the categories to train the algorithm on how to detect adversarial bots.

We also need to train a new algorithm ($A_7$) how to use modern alternatives such as Samsara (a Scala-backed DSL language that allows for in-memory and algebraic operations) and how to use Mahout to perform clustering, classification, and batch-based collaborative filtering.

*Table 4: Conceptual framework of algorithms as preventative solutions – corresponding with the cyber-risk forecasts*

| AI algorithms as preventative solutions for rogue AI systems | A |
|---|---|
| Solution 1: Synthesise new and emerging forms of data and use to develop more efficient algorithms**.** Then apply the algorithms to test the efficiency and power consumption while conducting predictive, dynamic, real-time quantitative risk analytics. | $A_1$ |
| Solution 2: Benchmarking emerging and unexpected risks from adversarial AI automating attacks. Develop AI based on compact representations, that can operate with lower memory requirements. | $A_2$ |
| Solution 3: Forecasting the most likely path to developing more efficient AI algorithms. | $A_3$ |

---

[4] https://www.qubit.com/

| | |
|---|---|
| Solution 4: Validation of the security readiness of AI systems: design for self-adapting AI systems compromised in a cyber-attack e.g., AI-driven bot launching an automated attack. | $A_4$ |
| Solution 5: Design for dynamic and self-adapting predictive (real-time) analytics of risks, i.e., forecasting cyber risk from AI. | $A_5$ |
| Solution 6: Construct tools and mechanisms for preventing bias in AI algorithms e.g., use of less biased/more inclusive data. Forecasting the risk and effect of catastrophic and existential events e.g., triggered by adversarial AI cyber-attack in combination global war, terrorist attacks. | $A_6$ |

To prove the effectiveness of the proposed conceptual framework, we pursued verification and comparison of results from established cyber security frameworks, to evaluate the performance of the proposed framework. The proposed conceptual framework (Table 4) is compliant with 'NVIDIA Morpheus Open AI Framework for Cybersecurity'[i]. This AI framework enables the development of AI pipelines for filtering, processing, and classifying large volumes of real-time data. The main advantages for verification and comparison of results from the conceptual framework (in Table 4) is the availability of pre-trained AI models with developer kits in AWS and Red Hat. In addition, the conceptual framework (in Table 4) is also compliant with the CEPS Task Force Report on Artificial Intelligence and Cybersecurity[ii]. This confirms further the validity, verification and comparison of results that are presented throughout this article. The verification and comparison of results in this article is grounded upon the recommendations from ENISA[iii], which state that: *'Before considering using AI as a tool to support cybersecurity, it is essential to understand what needs to be secured and to develop specific security measures to ensure that AI itself is secure and trustworthy.'*. In other words, while pursuing validity of results in terms of effectiveness of the AI framework in securing a system, we also considered that AI in-itself presents a risk to the evolving systems. Significant considerations have been placed on the NIST's efforts on AI and cybersecurity research[iv] and we hope this work would contribute to the future advancement of the NIST - AI risk management framework[v]. This article is presented in timely manner to the NIST request for public comments (a concept paper - by January 25, 2022) to help guide development of the AI Risk Management Framework.

# 7   Discussion

Evolving Systems are one kind of AI algorithms that own good performance of handling dynamic environments. They are effective solutions for cyber-attacks with dynamic characteristics. Evolving systems usually employ 'Fuzzy Logic', advanced 'Artificial Neural Networks' algorithms and hybrid approaches to derive optimised solutions for intelligent information analysis and visualisation [25]. Evolving systems emerge from the *'interaction and cooperation'* with adaptive structures, and *'derive decision patterns from stream data produced by dynamically changing environments'*, where assembling the system can be consistent of: *'rules, trees, neurons, and nodes of graphs'*, in relations to *'time-varying environments'* [26]. That definition of evolving systems was used throughout this article and used in combination with emerging literature on the state-of-the-art in applying AI algorithms in cybersecurity. Building upon recent literature on management of dynamic complex systems in cyber-attacks', this article expanded on existing efforts that *'characterize* *the process of self-regulation evaluating the system's resistance to cyber attacks'* [27]. Evolutionary computation techniques have already been used to identify parameters, functions, estimations, and optimisations - to improve the performance of empirical scientific computing with theory time-varying data [28].

To present the new algorithms to non-experts, we could firstly develop engaging content in different formats and translate the complex algorithms into visual stories e.g., concept diagrams (Figure 3).

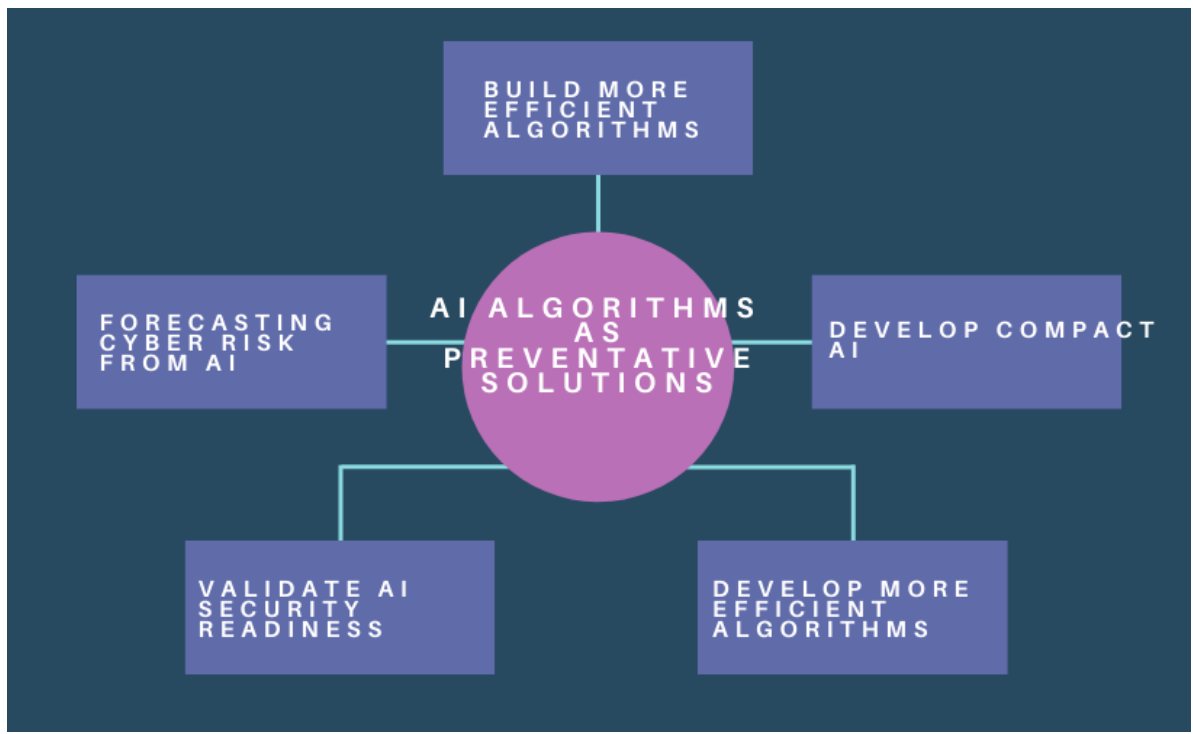

*Figure 3: Preventative solutions for rogue AI systems*

In Figure 3 we can see a visual representation of the proposed iterative methodology (repetition of processes to generate a sequence of preventative outcomes) for improved cybersecurity solutions resulting from the Super-Forecasting exercise in this article. One crucial point to note is that developing a solution is not the final security task, because adversarial AI will continue to evolve, and the solution needs to repeat the processes to ensure effectiveness from future AI risks.

Secondly, we need to present the new algorithms in a format that will be easy for professionals to understand e.g., 3D shaped versions of the algorithms. To disseminate the findings from and the algorithms, future researchers can use social media tools e.g., YouTube videos and target for global reach of non-technical audiences and to introduce the topic on applications of AI algorithms in cybersecurity.

In summary, by synthesising data on risk from AI, the study partially addressed the technological singularity hypothesis, where artificial super-intelligence leads to catastrophic events triggered by self-aware machines, with a focus on protecting the healthcare systems. The forecasts characterise synthesised data on cyber-risks from catastrophic events

Full reference:
Radanliev, Petar., De Roure, David., Maple, Carsten., and Ani, Uchenna, "Super-forecasting the 'technological singularity' risks from artificial intelligence," *Evol. Syst. 2022*, pp. 1–11, Jun. 2022, doi: 10.1007/S12530-022-09431-7, URL: https://link.springer.com/article/10.1007/s12530-022-09431-7.

triggered by AI attack to national critical infrastructure. 'The Singularity' is a technological singularity hypothesis, where artificial super-intelligence leads to self-aware machines. It is widely expected that when the 'The Singularity' occurs, it will 'abruptly trigger runaway technological growth, resulting in unfathomable changes to human civilization' [29]. While this can be seen as a distant future scenario by some researchers, in practice there are already solutions that resemble some of the forecasts described e.g., SingularityNET[vi]. This creates a strong rationale for further research and constructing solutions on how we can control such rogue AI machines.

# 8 Conclusion

This article presents a new framework for mitigating the risk from a 'technological singularity' event by using AI algorithms as preventative solutions for rogue AI systems. To construct the framework a set of forecasts are developed, based on the current knowledge of risks from artificial intelligence. Based on the forecasts, the new framework creates synergies between AI used to defend from cyber risk and defending from AI at the same time. The methodology applied in this article is based on a red-teaming approach assessing the risk of AI attack and derives forecast of AI rogue devices acting independently. This novelty of this research emerges in the form of risk forecasts synthesised in a framework for preventing risks from AI itself. The new framework analyses how we can secure a system, and how the system responds when compromised. Since not all systems can be secured, the emerging framework is grounded on enabling existing cyber-physical systems to continue operating when compromised. The presumptions in this article are based on the concept that any future 'superintelligence' would have intelligence much greater than the most intelligent human minds. Following this argument, *'it is difficult or impossible for present-day humans to predict what human beings' lives would be like in a post-singularity world'* [30]. This leaves very limited strategic options, but one that does remain available at present, it for humanity to continue doing what it has done for preventing global threats historically, and that is to form coalitions. With intelligence comes the ability for decision making, and as we can witness from human intelligence, two intelligent human beings can have two completely opposite perceptions of the world. It is likely that a future artificial 'superintelligence' would face similar decision-making challenges. If this presumption proves correct, then the mitigation strategy for a 'technological singularity' is the ability to form coalitions with the likeminded artificial 'superintelligence'.

## 8.1 Limitations

The personal perceptions of risk interact with data regulations, standards and policies need to be strongly integrated in the data analytics of the threat event frequencies (e.g., with a dynamic and self-adopting AI enhanced methodology). This will empower the design of mechanisms for predicting the magnitude through the control, analysis, distribution, and management of probabilistic data. The development of such research will enable deeper understanding of the impact of cyber risk at the edge. This would also define the baseline process for creating an updated risk/value impact model. Representing an advancement of the existing cyber risk assessments (e.g., developed on qualitative approaches, but with the added elements of AI, real-time intelligence, and dynamic risk analytics).

Full reference:

Radanliev, Petar., De Roure, David., Maple, Carsten., and Ani, Uchenna, "Super-forecasting the 'technological singularity' risks from artificial intelligence," *Evol. Syst. 2022*, pp. 1–11, Jun. 2022, doi: 10.1007/S12530-022-09431-7, URL: https://link.springer.com/article/10.1007/s12530-022-09431-7.

---

[i] https://developer.nvidia.com/morpheus-cybersecurity
[ii] https://www.ceps.eu/wp-content/uploads/2021/05/CEPS-TFR-Artificial-Intelligence-and-Cybersecurity.pdf
[iii] https://www.enisa.europa.eu/topics/iot-and-smart-infrastructures/artificial_intelligence
[iv] https://www.nist.gov/artificial-intelligence
[v] https://www.nist.gov/itl/ai-risk-management-framework
[vi] https://singularitynet.io/#